\documentclass[journal=ancac3,manuscript=article]{achemso}
\usepackage{amsmath}
\usepackage{graphicx}
\usepackage{xcolor}
\usepackage{hyperref}
\usepackage{physics}
\usepackage[version=4]{mhchem}
\usepackage[nameinlink,noabbrev,capitalize]{cleveref}
\usepackage[subpreambles=true]{standalone}
\usepackage{import}
\usepackage{pdfpages}

\title{Interfacial Energy Gradients Drive Coalescence of Supported Nanoparticles}
\author{Cheng-Yu Chen}
\affiliation{Department of Materials Science and Engineering, University of Pennsylvania, Philadelphia, PA 19104}
\author{Duncan Burns}
\affiliation{Department of Materials Science and Engineering, Northwestern University, Evanston, IL 60208}
\author{Peter W. Voorhees}
\affiliation{Department of Materials Science and Engineering, Northwestern University, Evanston, IL 60208}
\author{Eric A. Stach}
\affiliation{Department of Materials Science and Engineering, University of Pennsylvania, Philadelphia, PA 19104}
\email{stach@seas.upenn.edu}

\begin{document}

\begin{abstract}

Understanding and controlling nanoparticle coalescence is crucial for applications ranging from catalysis to nanodevice fabrication, yet the behavior of nanoparticles on dynamically evolving, heterogeneous substrates remains poorly understood.
Here, we employ \textit{in situ} transmission electron microscopy to investigate platinum (Pt) nanoparticle dynamics on silicon nitride (SiN$_x$) substrates where localized crystalline silicon (Si) nanodomains are deliberately formed via electron beam irradiation at $800^\circ$C.
We observe that Pt nanoparticles in contact with these Si pads transform into a more mobile platinum silicide (\ce{Pt3Si}) phase.
Strikingly, these \ce{Pt3Si} nanoparticles exhibit pronounced directional migration away from the Si pads, driven by interfacial energy gradients, rather than undergoing stochastic Brownian motion.
This directed movement fundamentally dictates coalescence pathways, leading to either enhanced sintering when particles are channeled together or inhibited coalescence when Si pads act as repulsive barriers.
Our findings reveal that local substrate chemistry and the resulting interfacial energy landscapes can dominate over initial particle size or proximity in controlling solid-state nanoparticle migration and assembly.
This work provides   insights into how substrate heterogeneity can be used to direct nanoparticle behavior, challenging conventional coalescence models and offering pathways for the rational design of supported nanomaterials.

    
\end{abstract}

\maketitle

\section{Introduction}

Supported nanoparticles (NPs) are crucial in diverse fields, from heterogeneous catalysis to nanoelectronics, owing to their high surface-area-to-volume ratios and unique size-dependent properties \cite{Cuenya2010}. However, their long-term performance is often undermined by thermal coarsening, or sintering, which leads to a loss of active surface area and functional characteristics \cite{Liu2018}. Classical sintering mechanisms for supported NPs primarily include Ostwald Ripening (OR), involving atomic diffusion, and Particle Migration and Coalescence (PMC), where entire particles diffuse and merge \cite{Hansen2013}. These processes are significantly influenced by factors like temperature, atmosphere, and nanoparticle-support interactions, with strong metal-support interactions (SMSI) potentially leading to complex behaviors such as phase transformations \cite{Luo2022, Kim2010}.

Beyond these thermally driven random processes, the directed motion of matter induced by interfacial energy gradients offers a powerful, yet relatively unexplored, avenue for deterministic control at the nanoscale \cite{Anderson1989}. The fundamental principle dictates that systems may evolve to minimize total interfacial free energy; consequently, spatial variations in this energy can manifest as a net thermodynamic force. This concept is well-established and has been extensively leveraged in manipulating \textit{liquid droplets} \cite{Brochard1989, Chaudhury1992, Liu2017}. Gradients in surface wettability or chemical potential, for instance, have been shown to compel highly directional droplet motion, a phenomenon that underpins innovations in microfluidics, coating technologies, and responsive surfaces \cite{Sun2005, Dai2020, Gulfam2022}. Similarly, a gradient in interfacial energy also gives rise to diffusion through the bulk or along the interface that leads to the motion of solid particles in a solid matrix  \cite{zhang2022migration}. The rich body of research into liquid systems has provided a deep understanding of how their behavior can be precisely controlled by engineering the interfacial energy landscape.

In stark contrast, the directed migration of \textit{solid-state nanoparticles} driven by such interfacial energy gradients—and its subsequent influence on their coalescence or assembly—remains a significantly less explored frontier. While solid NPs are, in principle, subject to the same thermodynamic forces, their characteristically lower atomic mobility at temperatures below their melting point often leads to an assumption that their movement is dominated by stochastic Brownian-like motion or is otherwise severely kinetically hindered \cite{Horwath2021}. Thus, a knowledge gap exists in understanding if and how deliberately engineered or spontaneously formed interfacial energy landscapes can overcome these kinetic barriers to induce deterministic, long-range transport and dictate the organization of solid nanoparticles. Bridging this gap is essential not only for designing next-generation nanomaterials with enhanced thermal stability but also for unlocking novel strategies for the directed assembly of solid-state nanostructures with precisely controlled architectures and functionalities.

This study directly addresses this knowledge gap by investigating the dynamic behavior of platinum (Pt) nanoparticles on silicon nitride ($\text{SiN}_x$) substrates where localized crystalline silicon (Si) nanodomains are formed \textit{in situ} via electron beam irradiation at $800~^\circ\text{C}$. We demonstrate that Pt nanoparticles, upon contact with these Si pads, transform into a more mobile platinum silicide (\ce{Pt3Si}) phase. Crucially, these solid \ce{Pt3Si} nanoparticles exhibit pronounced directional migration away from the Si pads. This directed movement, driven by interfacial energy gradients between the \ce{Pt3Si} phase, the Si pads, and the surrounding $\text{SiN}_x$ substrate, fundamentally dictates subsequent coalescence pathways. Depending on the spatial arrangement of the Si pads, these interactions can either enhance sintering by channeling nanoparticles together or inhibit coalescence by creating repulsive barriers.

Through a combination of \textit{in situ} transmission electron microscopy (TEM) observations, which allow real-time tracking of individual nanoparticle trajectories and phase transformations, and phase-field modeling, we elucidate the mechanisms governing this directed solid-state migration. Our findings reveal that local substrate chemistry and the dynamically evolving interfacial energy landscape can dominate over initial particle size or proximity in controlling nanoparticle behavior, challenging conventional coalescence models that primarily rely on statistical encounters. This work provides direct mechanistic insights into how \textit{in situ} generated substrate heterogeneity can be used to deterministically guide solid nanoparticle motion and assembly, offering pathways for the rational design of supported nanomaterials with tailored properties and stability.

\section{Results and Discussion}

To investigate how substrate heterogeneity influences nanoparticle behavior at elevated temperatures, we deliberately induce surface modifications on silicon nitride substrates by subjecting them to sustained electron beam exposure at $800^\circ$C. We find that this treatment consistently results in the formation of crystalline silicon nanodomains, as confirmed by selected area electron diffraction (SAED) and high-resolution transmission electron microscopy (HRTEM) (Supplemental Figure~1). These nanocrystalline Si features nucleate and grow under prolonged beam exposure at $800^\circ$C, introducing randomly distributed spatial variations in interfacial energy across the substrate surface.

\begin{figure}[htbp]
    \centering
    \includegraphics[width=0.75\textwidth]{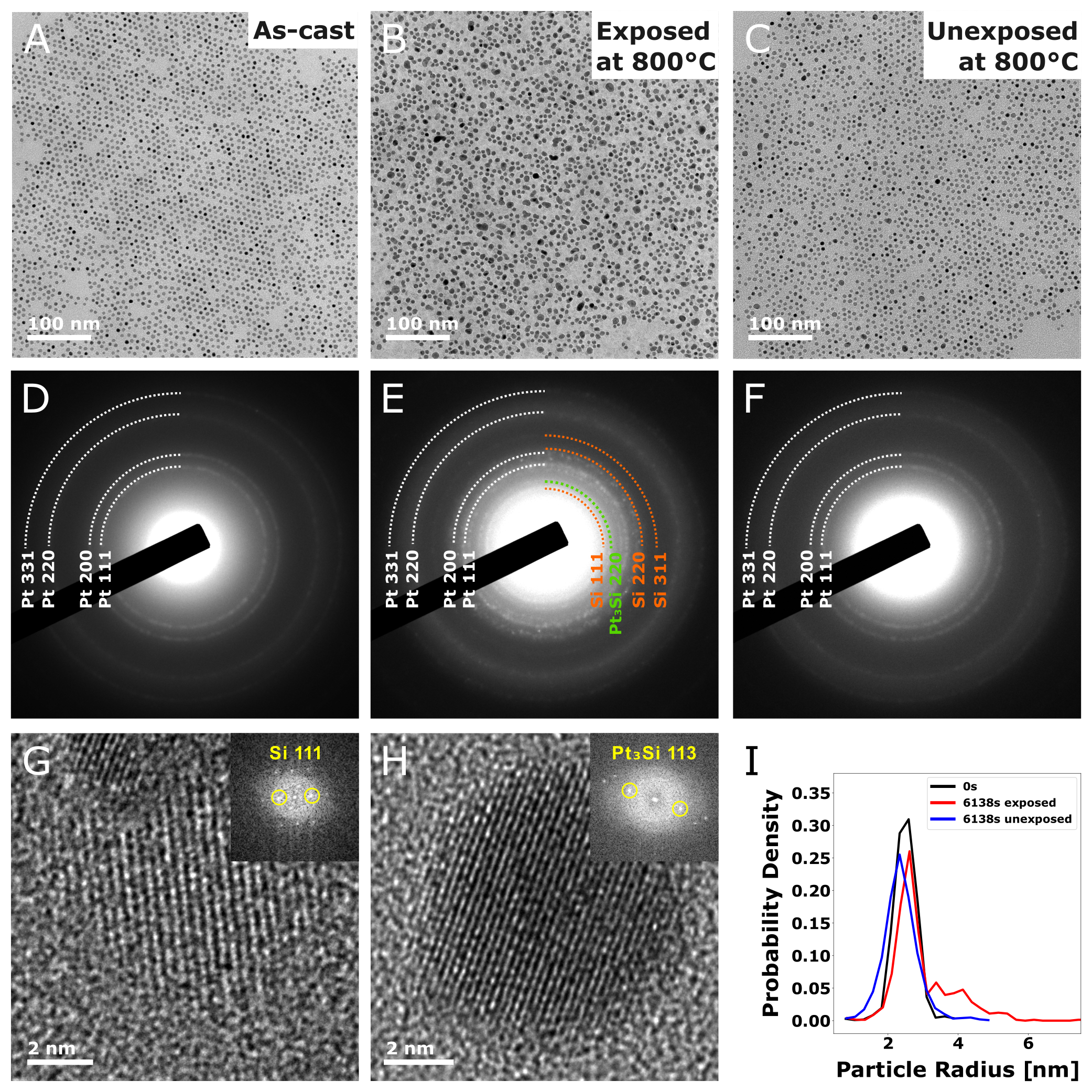}
    \caption{Observation of surface heterogeneity enhanced coalescence. (A, D) As-cast Pt nanoparticles and SAED pattern showing Pt FCC structure. (B, E) E-beam exposed region after heating at 800 °C and its SAED pattern with additional reflections of monoclinic Pt$_3$Si and Si. (C, F) E-beam unexposed region after heating at 800 °C and its SAED pattern, showing less coalescence and the retention of the FCC structure. (G) HRTEM image of a sintered particle from the exposed region and its FFT, showing lattice fringes consistent with the Pt$_3$Si structure. (H) HRTEM image and its FFT showing the formation of a Si nanocrystal on the substrate. (I) Particle size distributions of the unexposed region compared with the exposed region over time.}
    \label{fig:beam_effect}
\end{figure}

To study how such engineered heterogeneity affects particle behavior, we deposit platinum nanoparticles onto silicon nitride heating chips and observed their structural evolution under the same electron irradiation and heating conditions. Our model catalyst system consists of monolayer self-assembled Pt nanoparticles supported on a silicon nitride membrane. The particles exhibit a monodisperse size distribution with an average radius of $R \approx 2.5$~nm. The Pt nanoparticles, coated with dendrimer ligands to promote defined spacing during self-assembly, were synthesized via solvothermal methods~\cite{Yang2023} and drop-cast onto Hummingbird Scientific heating chips~\cite{2024}. Self-assembly ensured sufficient interparticle separation to prevent overlapping during TEM imaging, facilitating particle labeling and tracking. Ligand removal was performed using low-power oxygen plasma cleaning~\cite{Ristau2009, Cortie2013}, minimizing sintering temperature suppression due to organic coatings. Figure~\ref{fig:beam_effect}A shows a representative nanoparticle array following plasma treatment.

The particle array is then heated to $800^\circ$C within the TEM for more than 1.5 hours. Figures~\ref{fig:beam_effect}A--C compare the same region before heating, after beam exposure, and in an adjacent unexposed region. After heating, the initially uniform monolayer transforms into a non-uniform distribution containing several larger sintered particles. SAED patterns reveals distinct structural differences: the as-cast sample exhibited diffraction rings consistent with face-centered cubic (FCC) Pt (Figure~\ref{fig:beam_effect}D), while the beam-exposed regions developed additional rings corresponding to crystalline Si and monoclinic \ce{Pt3Si} (Figure~\ref{fig:beam_effect}E)~\cite{Gohle1964, Ram1978}. In contrast, the unexposed region retains its original FCC structure with no new phases (Figure~\ref{fig:beam_effect}F). In contrast, the unexposed region retains its original FCC structure with no new phases (Figure~\ref{fig:beam_effect}F). While coalescence is significantly less pronounced in these areas, slight rearrangements of nanoparticles are still observed (Figure~\ref{fig:beam_effect}C), suggesting that weak attractive forces exist even in the absence of beam-induced transformation. Given the low experimental temperature relative to the melting point of platinum ($T_m \approx 1768^\circ$C), such mobility is limited and likely governed by van der Waals interactions.

These observations highlight the critical role of beam-induced effects in enabling phase transformation and enhanced particle mobility, prompting further investigation into the structural changes underlying this behavior. HRTEM imaging then further confirms the structural transformation. Lattice-resolved imaging of particles from the beam-exposed region reveals lattice spacings consistent with crystalline Si (Figure~\ref{fig:beam_effect}G) and \ce{Pt3Si} (Figure~\ref{fig:beam_effect}H). Si islands were identified on the silicon nitride window, situated both in contact with Pt nanoparticles and in particle-free areas (Supplemental Figure 2). While Si nucleation occurred relatively randomly across the irradiated area of the substrate, the particle-substrate interface could have functioned as a heterogeneous nucleation site, marginally favoring localized Si formation.

Previous studies have reported that \ce{Pt3Si} forms in \ce{Pt/SiO2} systems under hydrogen environments at temperatures above 600~$^\circ$C, often accompanied by substantial sintering~\cite{Lamber1987, Wang2003, Behafarid2014}. These transformations typically proceed via a cubic \ce{Pt3Si} intermediate~\cite{Lamber1991}. In contrast, the formation of metal-rich \ce{Pt3Si} rather than more Si-rich phases such as \ce{Pt2Si} or \ce{PtSi} suggests that silicon availability is locally limited in our system~\cite{Canali1977, Canali1979, Ottaviani1978}. The absence of silicide formation in the unexposed region—despite identical temperature and time—confirms that the electron-beam-induced formation of Si islands is a prerequisite for \ce{Pt3Si} nucleation on amorphous silicon nitride substrates.

To further validate the temperature-dependent onset of silicide and silicon crystallization, we perform \textit{in situ} electron diffraction as a function of temperature (Supplemental Figure~4). SAED patterns were recorded at 50~$^\circ$C intervals from 300 to 800~$^\circ$C, followed by a time-resolved series at 800~$^\circ$C. Radially integrated intensity profiles show that the \ce{Pt3Si} (220) reflection emerges between 700 and 750~$^\circ$C and intensifies over time at 800~$^\circ$C, while the Si (220) reflection increases steadily under beam exposure. The stability of the Pt (331) peak confirms that only a fraction of Pt nanoparticles undergo phase transformation. These findings indicate that the formation of \ce{Pt3Si} requires both high temperature and contact with beam-induced Si pads, and that these pads act as localized, rather than global, sources of silicon.

Comparison of particle size distributions between exposed and unexposed regions (Figure~\ref{fig:beam_effect}I) further highlights differences in sintering behavior. In the unexposed regions, the particle size distribution retains a higher population of smaller particles near the initial distribution peak, suggesting that coarsening occurs uniformly but slowly, primarily via Ostwald ripening. Conversely, in areas exposed to the electron beam, the formation of Si pads and the subsequent transformation of Pt to a more mobile \ce{Pt3Si} phase  promote increased particle mobility; this leads to significant coarsening via Particle Migration and Coalescence (PMC), as evidenced by the distinct tail of larger particles in the size distribution. These results demonstrate that beam-induced substrate heterogeneity alters interfacial energy landscapes, enabling distinct coalescence behaviors between otherwise identical regions of the sample.

\newpage

To further understand the dynamics of coalescence driven by surface heterogeneity, we conduct automated image analysis on a 1.5-hour \textit{in situ} image sequence collected at 800~$^\circ$C (Supplemental Movie~1). Image segmentation is performed using an unsupervised U-Net-based model to distinguish foreground particles from the background~\cite{Horwath2020, Vyas2022}. Post-processing enabled identification and labeling of individual nanoparticles, resulting in a quantitative dataset of particle sizes and positions over time. Further details of the image analysis pipeline are described in the \hyperref[sec_methods]{Methods} section.

\begin{figure}[htbp]
    \centering
    \includegraphics[width=\textwidth]{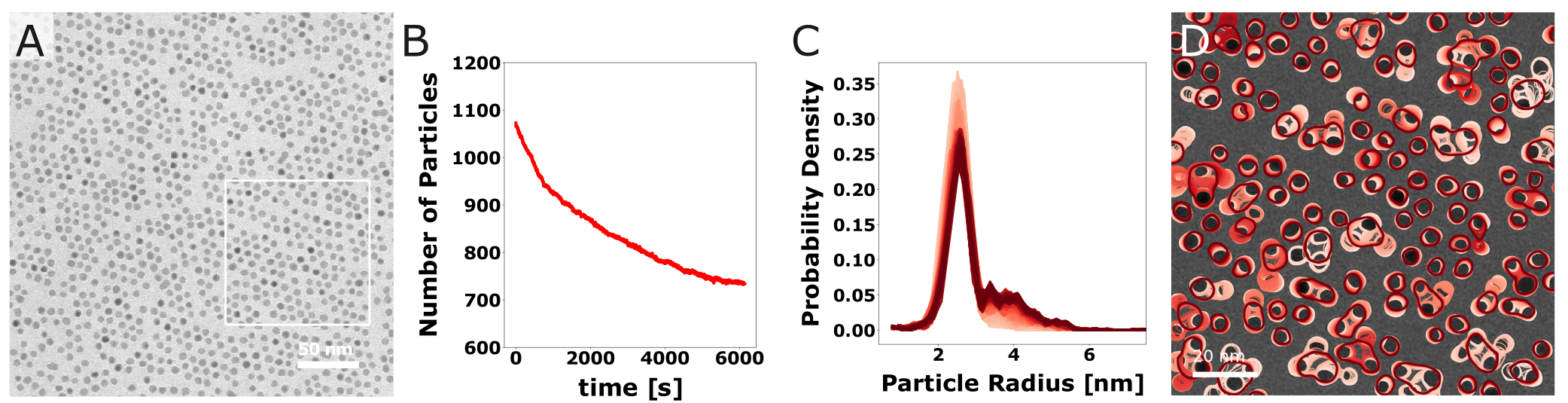}
    \caption{Overview of statistical data showing selective coalescence. (A) The as-deposited array of platinum nanoparticles. (B) The number of particles as a function of time. (C) The evolution of the particle size distribution over time, with the shade of red darkening as time progresses. (D) Colored outlines represent particle boundaries, with time progressing from light to dark red. The particles shown are within the boxed regions in (A).}
    \label{fig:insitu_analysis}
\end{figure}

An overview of the statistical data is presented in Figure~\ref{fig:insitu_analysis}. The color-coded particle outlines reveal substantial changes in nanoparticle size and spatial distribution (Figure~\ref{fig:insitu_analysis}D). Although most particles remain stationary and unchanged in size, a distinct subpopulation undergoes directional migration and coalescence. This behavior leads to a gradual decrease in particle number (Figure~\ref{fig:insitu_analysis}B) and a concurrent broadening of the size distribution with a growing tail of larger particles (Figure~\ref{fig:insitu_analysis}C). These observations are consistent with the limited spatial coverage of Si pads formed via electron beam irradiation, which locally promote enhanced mobility and coalescence.

Closer inspection of particle motion reveals that coalescence is associated with long-range shifts in the center-of-mass positions of individual particles, indicating directional migration rather than stochastic Brownian motion. This behavior contrasts with conventional surface diffusion-limited coarsening mechanisms~know\cite{Hansen2013, Horwath2021}. Moreover, we observe that some neighboring particles did not coalesce, despite exhibiting comparable sizes and separation distances to those that did. To evaluate whether initial size or proximity influenced coalescence behavior, we map the positions of large particles and closely spaced pairs relative to the coalesced population. As shown in Supplemental Figure~5, no strong correlation was found between these metrics and the observed coalescence, suggesting that local substrate chemistry, rather than geometric factors, governs the dynamics.

Because the field of view used in the \textit{in situ} video acquisition prioritized wide-area coverage, it does not allow direct observation of individual Si pads during particle motion. To elucidate the interactions between particles and Si pads, we correlate high-resolution TEM images acquired after heating with the same regions observed during the dynamic video sequence (Figure~\ref{fig:HRinspection}; additional examples in Supplemental Figures~6 and 7).

In Figure~\ref{fig:HRinspection}A, a Si pad (pseudocolored in red) is  in direct contact with the edge of a large sintered particle. The crystalline Si region was identified using geometric phase analysis of a selected FFT peak~\cite{Jeon2021}. Corresponding \textit{in situ} frames show that this sintered particle resulted from the merging of two initially separated nanoparticles. Over the first 600 seconds, the particles draw closer and fuse. Using the fixed reference arrows across frames, we track the center-of-mass trajectory and found that the upper-left portion of the particle remained nearly fixed, while the lower-right portion migrated upward and to the left—away from the Si pad—throughout the coalescence process.

Additional insights are gained by examining the behavior of nearby non-sintered particles. Center-of-mass trajectories plotted in the final frame of Figure~\ref{fig:HRinspection}A show that one neighboring particle (purple) migrated away from the large sintered particle, presumably repelled by the Si pad. In contrast, the other two neighboring particles (red and orange), which had no Si pad between them and the sintered particle, moved slightly closer. However, no coalescence occurred over the full 1.5-hour experiment. This suggests that van der Waals attraction between particles in Si-free regions is weak, consistent with the reduced sintering observed in the unexposed control region of Figure~\ref{fig:beam_effect}C.

A second example is shown in Figure~\ref{fig:HRinspection}B, where a Si pad (pseudocolored in red) is located between two non-sintered particles. Time-resolved images show that these particles move closer together until 900 seconds, reaching a minimum separation. A red bar, drawn between the two particles in each frame, serves as a visual reference for their evolving interparticle distance—initially shortening, then lengthening slightly, and finally stabilizing. The spacing remained approximately constant at $\sim$1~nm through the end of the experiment at 6138 seconds. The initial approach up to 900 seconds is likely driven by van der Waals attraction between the two particles. Beyond that point, the particles exhibit shape adjustments and lateral shifts, likely reflecting complex multiparticle interactions and constrained motion along the edge of the intervening Si pad. After cooling, HRTEM imaging revealed a further increase in spacing to $\sim$2~nm at room temperature. This behavior suggests that the attractive van der Waals interparticle forces were balanced by repulsion from the intervening Si pad, resulting in an equilibrium separation. These competing interactions limited further approach or coalescence, reinforcing the conclusion that local Si pads modulate interparticle mobility and coarsening dynamics by altering interfacial energy landscapes.

Furthermore, long-range directional motion of a non-sintered particle was observed in Figure~\ref{fig:HRinspection}B. A nearby Si pad, pseudocolored in yellow in the HRTEM image, appears to influence its trajectory. By using the yellow arrow as a fixed spatial reference (initially pointing to the center-top edge of the particle), we observe that the particle consistently migrates downward and to the right over time—away from the Si pad. The center-of-mass trajectory confirms significantly greater displacement compared to surrounding particles, suggesting that this directional migration is related to the presence of the Si domain.

An another notable aspect of these Si pad-induced particle dynamics, evident from the detailed analyses in Figure~\ref{fig:HRinspection} and Supplemental Figures~6 and 7, is the variability in the initiation times for directed particle movement. Different nanoparticles, or groups of nanoparticles, commence their Si pad-driven migration at different stages of the \textit{in situ} experiment. This observation strongly suggests that the formation of Si pads is not a singular event occurring uniformly across the substrate at the outset of heating and beam exposure, but rather is a progressive process. Such continuous nucleation and growth of Si nanodomains throughout the experiment are consistent with the gradually increasing intensity of the crystalline Si diffraction peaks observed in the time-resolved SAED data (Supplemental Figure~1 and Supplemental Figure~4). Furthermore, if the interfacial energy gradient from a Si pad drives a \ce{Pt3Si} nanoparticle away, the characteristic length scale of such a directed migration event is likely related to the dimensions of the Si pad exerting the influence. The persistent decrease in the total number of nanoparticles and increase in the number of larger particles observed throughout the 1.5-hour \textit{in situ} heating period (Figure~\ref{fig:insitu_analysis}B), indicative of ongoing coalescence, further supports the hypothesis of continuous Si pad formation. New Si pads forming over time would continually activate previously stationary \ce{Pt3Si} nanoparticles or alter the trajectories of already mobile ones, sustaining the coalescence activity.

Taken together, these detailed \textit{in situ} observations, considering both the spatial interactions and their temporal evolution, unequivocally demonstrate that Pt-based nanoparticles exhibit a strong tendency to migrate away from these progressively forming, beam-induced Si pads at $800^\circ$C. Depending on the evolving spatial arrangement of these Si pads relative to the nanoparticles, this surface energy gradient can either promote coalescence (by directing particles towards each other in Si-free zones) or hinder it (by creating repulsive barriers between particles).
The observed differences in both the magnitude of displacement and the timescale of motion strongly suggest that Si pad-induced repulsion is a dominant force, capable of overriding weaker, attractive interparticle forces under our experimental conditions. These results firmly establish the role of local, dynamically generated substrate heterogeneity as a primary driver of the observed directional migration and selective coalescence behavior in this supported nanoparticle system.

\begin{figure}[htbp]
    \centering
    \includegraphics[width=\textwidth]{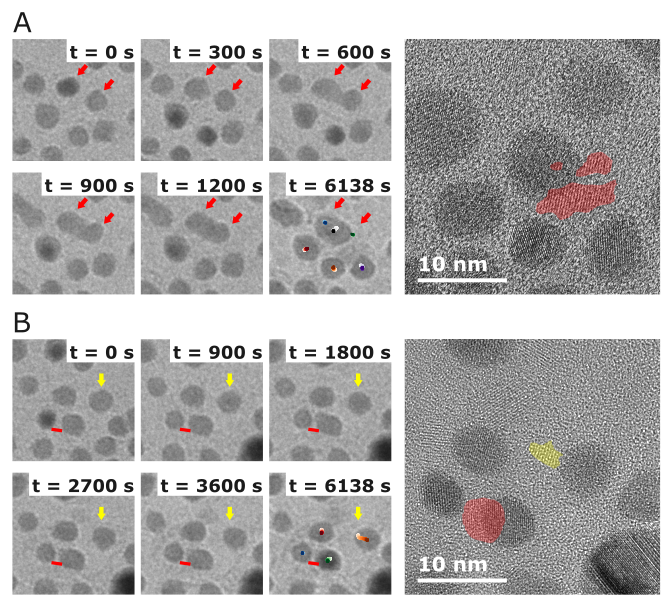}
    \caption{Time sequence analysis showing particle-substrate interaction. (A) A Si pad nearby a sintered particle located in the same direction from which one particle moved away. (B) A Si pad located between two non-sintered particles, preventing them from approaching each other. Another Si pad is located along the trajectory of a moving particle. The motions of the center of mass of particles are plotted on the last frames of (A) and (B), with the shade of color darkening as time progresses.}
    \label{fig:HRinspection}
\end{figure}
\newpage

To better understand the mechanisms driving this dynamic response, we perform phase-field modeling supported by analytical scaling arguments. These simulations aim to clarify how interfacial energy gradients and local substrate interactions contribute to nanoparticle migration, complementing the experimental findings presented in Figures~\ref{fig:beam_effect}--\ref{fig:HRinspection}.

In our case, Pt nanoparticles are initially deposited on an amorphous silicon nitride window and exhibit limited mobility due to the low experimental temperature relative to platinum’s melting point ($T_m \approx 1768^\circ$C). Interparticle attraction is presumed to occur primarily through van der Waals forces. However, under sustained heating and electron beam irradiation at $800^\circ$C, Si-rich domains nucleate on the substrate surface. At sites of contact between Pt particles and these Si pads, silicon diffuses into the particles, leading to the formation of monoclinic \ce{Pt3Si}. The melting temperature of \ce{Pt3Si} of 874$^\circ$C is much lower than the Pt melting point of 1769$^\circ$C, implying that there can be significant bulk and surface diffusion at 800$^\circ$C and, potentially particle  mobility. Once transformed, these nanoparticles are therfore more responsive to local driving forces and tend to migrate away from Si pads, ultimately coalescing on the surrounding silicon nitride surface.


\begin{figure}[htbp]
   \centering
    \includegraphics[width=\textwidth]{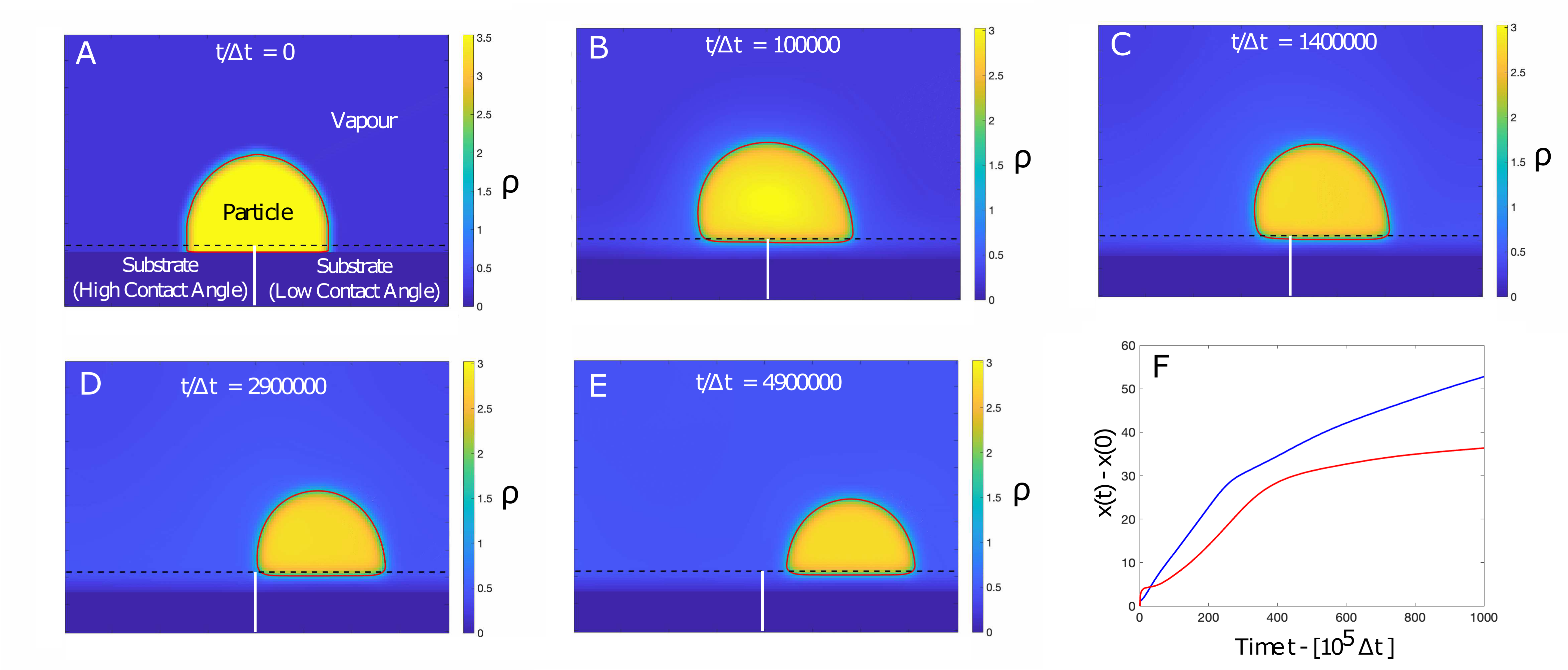}
    \caption{Two-dimensional phase field simulation of a nanoparticle cross-section migrating on a chemically heterogeneous substrate (held fixed in time). Left and right-hand side particle-substrate interface energies are set such that the equilibrium contact angle of $\theta_{Y}\approx140^\circ$ (gradient energy coefficient $\epsilon_{\text{left}} = 1000$) and $\theta_{Y}\approx130^\circ$ (gradient energy coefficient $\epsilon_{\text{right}} = 1500$) for the left and right respectively. The colormap is representative of the dimensionless local material density, with the diffuse interface between particle-vapor a result of the model parameters chosen. The $\phi=0.5$ contour is demarcated by a red line distinguishing between particle and vapor. (A) At $t/\Delta t = 0$, the particle is initialized with an out of equilibrium contact angle. Here $\Delta t$ denotes a timestep increment set to capture bulk diffusion. (B) By $t/\Delta t = 100000$, the two contact angles evolve toward equilibrium, creating asymmetric curvature around the particle. (C) At $t/\Delta t = 1400000$, relaxation of interfacial curvature drives the particle toward the lower contact angle side. (D) At $t/\Delta t = 2900000$ particle reaches the lower contact angle side, but retains the asymmetric curvature. (E) Curvature relaxation causes particle to migrate further into the lower contact angle substrate side. (F) An illustration of the left (blue) and right (red) contact point relative to their initial positions ($x(t) - x(t=0)$) as a function of time.}
    \label{fig:PhaseField}
\end{figure}
To illustrate nanoparticle migration as a result of substrate heterogeneity, we employ a recently developed phase field approach (see \hyperref[sec_methods]{Methods} for further details.). Herein an order parameter $\phi$ and dimensionless density $\rho$ are evolved concomitantly to capture both solid-vapor interface kinetics in addition to diffusion-limited motion. A simulation of how a nanoparticle cross-section responds to substrate heterogeneity is illustrated in Fig.~\ref{fig:PhaseField}. The nanoparticle is first initialized with an out-of-equilibrium contact angle where the substrate has differing interfacial energies on each side of the simulation domain. We make the assumption that the substrate is non-evolving. The particle first adjusts its contact angles to the equilibrium contact angles and subsequently migrates through bulk diffusion towards the side with the lower contact angle due to the non-axisymmetric particle-vapor interfacial curvature. This small contact angle difference leads to a near constant velocity particle migration, the particle curvature and particle-vapor surface area appear to remain constant. Once the three-phase contact region of the nanoparticle is within the region where of the substrate where the contact angle is lower, the particle slows down and the equilibrium constant curvature shape of the particle-vapor interface is reestablished.

The phase field model is motivated as an expansion about the solid-liquid-vapor triple point and captures effects of evaporative kinetics and bulk phase diffusion. As a result, in addition to curvature-induced migration, sublimation from the  particle to the vapor occurs, reducing the overall size of the particle. Furthermore, an artificially broad density profile is established near the solid-vapor interface as a result of the bulk solid free energy curvature. Sublimation effects are not expected for the platinum nanoparticles in the current study. Nonetheless, these effects may be important for systems in which the solid particles can sublimate \cite{Horwath2020,meng2021anomalous}. However, the natural tendency of motion towards the lower contact angle side occurs regardless of slow sublimation.


The phase field simulations show that particle migration can result from interfacial energy gradients. We now seek to use the particle migration morphology captured in the phase field simulations to identify why some nanoparticles with Si pads between them were attracted until a stable separation was reached. We postulate that there is a force balance between the force due to interfacial gradients moving the particles apart and van der Waals interactions that induce attraction. These analytics will further elucidate the dependence of the migration process on materials parameters. 
\begin{figure}[htbp]
   \centering
    \includegraphics[width=\textwidth]{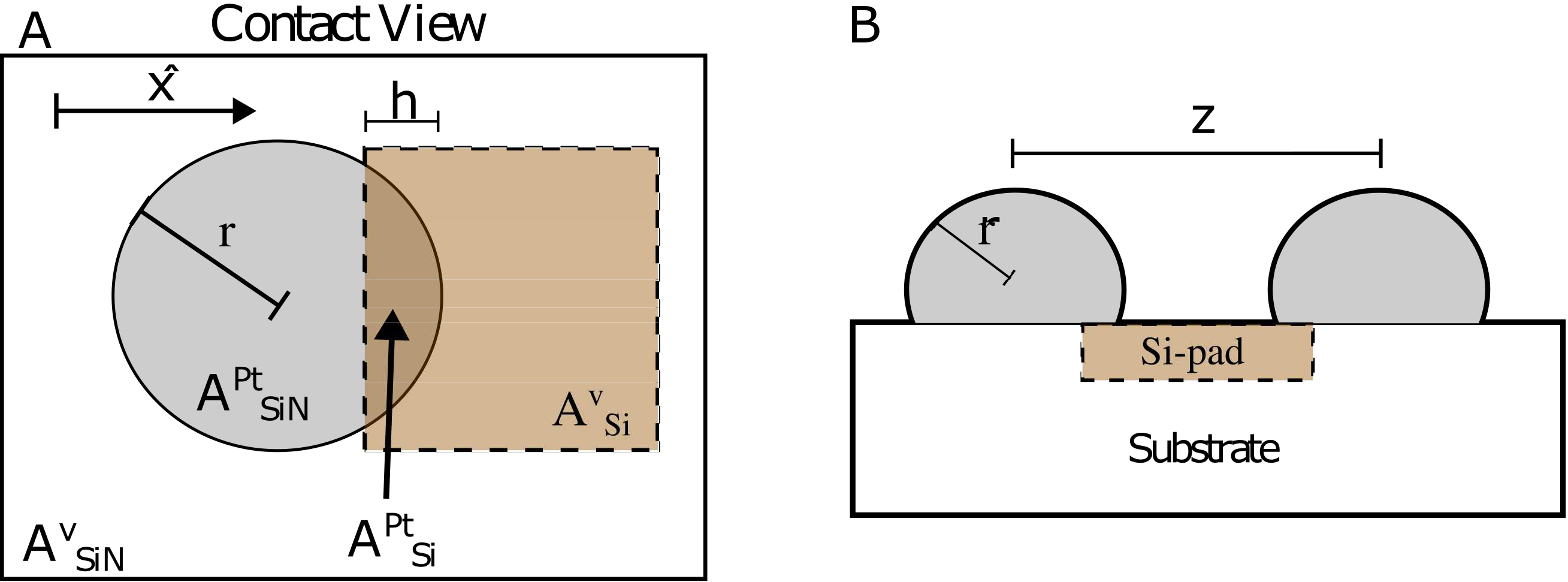}
    \caption{(A) View in the plane of the the particle-substrate contact. The gray region denotes the location of the particle. Circular segment geometry of particle contact with silicon-pad is highlighted by the region of color overlap, $A^{\ce{Pt}}_{\ce{Si}}$. (B) A side-view of two particles interacting with center of mass separation, $z$. Particles are assumed to have the same radius as the substrate particle contact, $r$.}
    \label{fig:analytic_schematic}
\end{figure}

Assuming equilibrium at the trijunctions and isotropic interfacial energies, the contact angles ($\theta_Y$) satisfy,
\begin{equation}\label{eq:Youngs_Law}
    \gamma_{\text{pt-vap}}\cos \theta_Y = \gamma_{\text{pt-sub}} - \gamma_{\text{vap-sub}},
\end{equation}
where $\gamma_{\text{pt-vap}}$ denotes the particle-vapor (in our case particle-vacuum) surface energy, $\gamma_{\text{pt-sub}}$, the particle-substrate surface energy, $\gamma_{\text{vap-sub}}$, the vapor-substrate energy. When the contact angle is given by Young's equation at all points along the trijunction, and particle-vapor interface has a constant curvature, the particles will be in equilibrium, and they will not migrate by interfacial energy gradients.

Turning our attention to the geometry in Fig.~\ref{fig:analytic_schematic}A,
a nanoparticle has contact area with a silicon pad, $A^{Pt}_{Si}$ and contact area with the silicon nitride substrate, $A^{Pt}_{SiN}$. Assuming a constant total nanoparticle contact area, $A_{Pt} = A^{Pt}_{Si} + A^{Pt}_{SiN}$, the configurational energy is
\begin{equation}
    E = \gamma^{Pt}_{SiN} (A_{Pt} - A^{Pt}_{Si}) + \gamma^{Pt}_{Si}A^{Pt}_{Si} + \gamma^{v}_{SiN}(A_{sub} - (A_{Pt} - A^{Pt}_{Si})) + \gamma^{v}_{Si}(A_{pad} - A^{Pt}_{Si}) + \gamma^{v}_{Pt}A^{v}_{Pt} + E_{\text{bulk}},
\end{equation}
where $A_{sub}$ is the total area of silicon nitride, $A_{pad}$ is the total area of a single Si pad. Meanwhile, $A^{v}_{Pt}$, is the surface area of the particle exposed to the vapor. $E_{\text{bulk}}$ corresponds to any energies that may be present in the particle volume. From the phase field simulations, the total areas appear to remain unchanged as the particle evolves. Furthermore, bulk energetic contributions are assumed to remain fixed in time. We note that when considering simultaneous sublimation, such an approximation may be invalid.

A configurational force can be determined, by comparing the energy gain when perturbing the particle center of mass further atop a silicon pad. With the use of Eq.~\ref{eq:Youngs_Law}, the configuration force is
\begin{equation}\label{eq:force}
    F_c = \gamma^{v}_{Pt}(\cos(\theta_{Y, Si}) - \cos(\theta_{Y, SiN}))\pdv{A^{Pt}_{Si}}{x} - \gamma^{v}_{Pt}\pdv{A^{v}_{Pt}}{x},
\end{equation}
with $x$ directed toward the silicon pad.
The equilibrium contact angles of particle on different substrate materials are denoted by $\theta_{Y, Si(SiN)}$. We highlight, that in three dimensions the contact line intersects the the discontinuity between substrates. As a result of the discontinuity particle-vapor surface area or particle-substrate contact surface area may vary with position. We make the assumption $\partial A^{v}_{Pt} / \partial x \rightarrow 0$, since during our phase field simulations, (see Fig.~\ref{fig:PhaseField}) $A^{v}_{Pt}$ is nearly constant when the center of mass motion is at steady state. While the curvature varies along the interface, as it must to induce motion, it shape does not change significantly and thus it is in a regime of near constant surface curvature and area. In other words, a change in the contact angle (and hence curvature if the contact area is held fixed) during motion may be captured through a non-zero $\partial A^{v}_{Pt} / \partial x$. This assumption may breakdown when the particle is smaller than the distance over which the interface energy changes, in which case the particle surface curvature may vary as the particle migrates. 

We note that Eq.~\ref{eq:force} highlights the preferred direction of motion. First assume that in the direction of $x$, $\partial A^{Pt}_{Si} / \partial x > 0$. Then if $\theta_{Y, Si} > \theta_{Y, SiN}$, $F<0$, and the direction of motion is unfavorable. Meanwhile the opposite occurs for $\theta_{Y, Si} < \theta_{Y, SiN}$. Particle migration is toward the side of lower equilibrium contact angle, consistent with the phase field simulation results.  

Different approaches were taken to estimate equilibrium contact angles of particles on silicon and silicon nitride surfaces. First, Pt nanoparticles and SiO$_{2}$ spheres were co-deposited on a silicon nitride TEM grid via drop-casting their mixed solution. The sample was heated under vacuum at 800 °C for 1 hour using a sealed quartz-tube method \cite{Roth1994}. Direct TEM observations of the heating transformation are shown in Supplemental Figure 8. Before heating, the contact angle between Pt particles and SiO$_2$ was between 115° and 120°. After vacuum heating at 800 °C, transformed Pt$_3$Si particles were observed on the surface of SiO$_2$ spheres, with contact angles ranging from 126° to 136°. Separately, based on prior reports that show that Pt$_3$Si particles embed into the concave crystalline Si surfaces \cite{Lin2015}, and using the contact angle relation between flat and spherical surfaces \cite{Wu2015}, the contact angle of Pt$_3$Si/crystalline Si was calculated to be approximately 140° to 150°. The contact angle differences between Pt$_3$Si/SiO$_2$ and Pt$_3$Si/crystalline Si highlight the interface energy differences between Si islands and the silicon nitride substrate. 

The solid-vapor interfacial energy of Si is $\gamma \approx 1$-$1.5$ $J/m^2$ \cite{Hara2005, Lu2005}, while Pt, $\gamma^{v}_{Pt}\approx 1.5$-$2.0 J/m^2$ \cite{Kim2018}. However, for Pt$_3$Si, no surface energy is known. We approximate the surface energy of Pt$_3$Si as bounded between that of Pt and Si, with a slight preference toward the Pt value, $\gamma^{v}_{Pt_3Si}\approx 1.7 J/m^2$. 

Using the geometry of Fig.~\ref{fig:analytic_schematic}A, $\partial A^{Pt}_{Si} / \partial x = \partial A^{Pt}_{Si} / \partial h$. Treating the overlap as a circular segment results in, $\partial A^{Pt}_{Si} / \partial h = 2\sqrt{r^2 - (r - h)^2}$. We can estimate a maximal interfacial energy gradient force that occurs when  $h=r$ with a substrate contact divided in half by silicon and silicon nitride for a $r=2nm$ as $F\approx 1$-$10 eV/nm$, for the two bounding cases of $\gamma^{v}_{Si_3N_4}$. The energy cost of a particle moving atop a silicon region is large, suggesting particles that straddle the Si-pad and SiN$_4$ will migrate off the pad. The analytics result is hence consistent with experiments and simulations.

We can now compare the configurational force due to interfacial energy differences against the pairwise van der Waals interactions. Considering Fig.~\ref{fig:analytic_schematic}B, we can determine the equilibrium distance distance between particles. Since the contact angles of the nanoparticles with the substrate is on the order of $130^\circ$, we approximate the nanoparticle as spherical bodies with radius, $r$. In which case, the Van der Waals force can be determined as,
\begin{equation}\label{eq:van_der_waals_force}
    F_V = \frac{-32 A r^3}{3z^3(z^2 - 4r^2)^2},
\end{equation}
with $A$ the Hamaker constant \cite{Hamaker1937}. Tolias {\it et al.} tabulated Hamaker constants for different metals, for platinum, $A \approx 450 zJ$ \cite{Tolias2020}. In Eq.~\ref{eq:van_der_waals_force}, $z$ denotes the center of mass separation between the two particles. We may alternatively use the particle separation $\xi = z - 2r$. The negative $F_V$ indicates an attractive force. For two (r = 2nm) platinum nanoparticles with a center of mass separation of $z=5nm$, the van der Waals force is $F_V\approx -0.19 eV/nm$.  

We then can compare the magnitudes of $F_c$ and $F_V$ following Fig.~\ref{fig:analytic_schematic}. Here we make the approximation that the particle radius and contact area radius are equal. We then assume that at a particle separation of $z_i=5nm$, there is no silicon overlap, $h=0$. Furthermore, we assume the circular segment of silicon pad increases in the direction of the van der Waals force, we thus let $h = z_i - z$.
Equating $F_c$ and $F_V$ with the lower estimate of $F_V$, gives an equilibrium center of mass separation of $z\approx5 nm$. Silicon pads between platinum silicide nanoparticles may therefore act as an effective barrier against sintering when particles are driven by van der Waals forces.

Ultimately, the sporadic nature of particle coalescence is attributed to the complex interplay of competing forces and transformations. We observed cases in which motion away from Si pads appeared to reinforce natural attraction between particles, accelerating sintering. Conversely, we also noted regions where Si islands were present between nanoparticles, yet no coalescence occurred. In these cases, the energy that follows from grain boundary formation likely inhibits coarsening. Thus, particle motion is governed by a balance of asymmetric surface energetics, repulsive interactions from Si pads, and attractive interparticle forces. The stochastic distribution of Si pads—determined by beam profile and substrate response—adds further complexity, yielding the seemingly erratic coarsening patterns observed experimentally.

\begin{figure}[htbp]
   \centering
    \includegraphics[width=\textwidth]{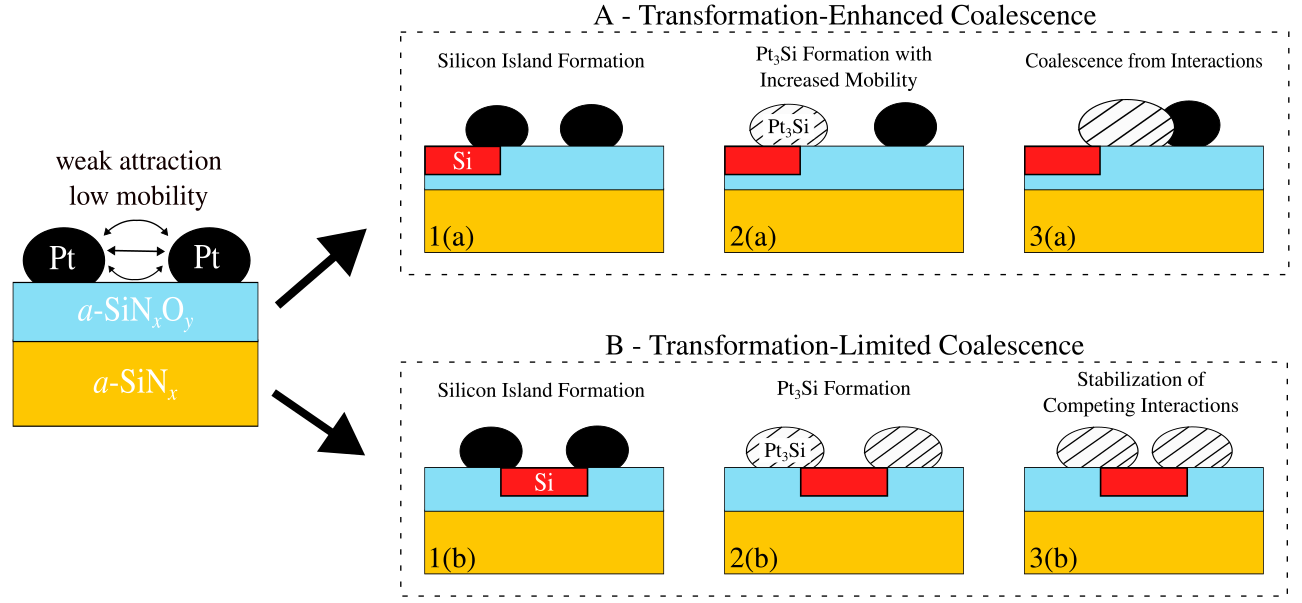}
    \caption{Schematic illustration of transformation-enhanced and transformation-limited coalescence behaviors in Pt nanoparticle systems. Initially, Pt nanoparticles supported on an amorphous silicon oxynitride (a-SiN$_x$O$_y$) surface exhibit weak attraction and low mobility. (A) In the transformation-enhanced pathway, silicon island formation (1a) facilitates the transformation of Pt into Pt$_3$Si with increased mobility (2a), leading to coalescence through reinforcing interactions (3a). (B) In the transformation-limited pathway, silicon islands also form (1b), and Pt$_3$Si transformation occurs (2b), but competing interactions between Pt$_3$Si particles and silicon islands stabilize the system, preventing coalescence (3b).}
    \label{fig:schematic}
\end{figure}

We thus identify two aspects of the dynamics of supported Pt nanoparticles at elevated temperatures. A schematic of the mechanism is provided in Figure~\ref{fig:schematic}. In particular, structural transformations of both the substrate and the nanoparticle material modify the sintering kinetics. As silicon pads nucleate due to the elevated temperatures and beam irradiation, Pt particles transform into more mobile platinum silicides. The increased mobility concomitantly increases susceptibility to driving forces thereby enhancing migration. In the preceding discussion, we have listed a number of different acting forces within our study and compared their relative magnitudes. The dominant process leading to migration is attributed to surface energy differences on each side of the particle due to substrate heterogeneity. Particle adherence to silicon regions is strongly unfavored. As a consequence, silicon pads can act to inhibit particle coalescence when silicon is present between neighboring nanoparticles, a situation of {\it limited coalescence}. On the contrary, when the silicon region is found on the one side of the \ce{Pt3Si} particles, migration occurs and coalescence is enhanced. Based on the significant overall increase in sintered particles observed in the exposed regions compared to the unexposed areas (Figure~\ref{fig:beam_effect}), this latter pathway—where Si pads drive migration that leads to {\it enhanced coalescence}—appears to be the more dominant outcome influencing the macroscopic coalescence behavior under these conditions. Strong correlations traditionally linked to interparticle separation may therefore be masked as a result of silicon pad formation. We suspect that for other unstable substrates types, nucleation may also mask spatial correlations. Future research may then allow exploitation of the effects.

\section{Conclusions}

In summary, this study has elucidated the impact of deliberately engineered substrate heterogeneity on the dynamic behavior of supported platinum nanoparticles at elevated temperatures.
Through \textit{in situ} transmission electron microscopy, we have demonstrated that sustained electron beam exposure induces the formation of crystalline silicon nanodomains on an amorphous silicon nitride substrate.
These Si pads act as localized sources of silicon, triggering the transformation of Pt nanoparticles into a more mobile \ce{Pt3Si} phase at $800^\circ$C, a temperature proximate to the Pt-Si eutectic.

Our key finding is that the subsequent migration and coalescence of these \ce{Pt3Si} nanoparticles are not governed by stochastic Brownian motion, but rather by deterministic, directional migration driven by interfacial energy gradients.
Specifically, \ce{Pt3Si} nanoparticles exhibit a distinct tendency to move \textit{away} from the Si pads.
This directed movement, dictated by the local interfacial energy landscape, fundamentally alters coalescence pathways.
Depending on the spatial arrangement of the Si pads, this interaction can either enhance coalescence by channeling nanoparticles together in Si-free regions or inhibit it by creating repulsive barriers between particles.
We find that this local substrate chemistry, rather than initial particle size or interparticle proximity, is the primary determinant of the observed selective coalescence patterns.

The significance of these findings is multifold. Firstly, they challenge the conventional understanding of nanoparticle coalescence on supported systems, revealing that directional migration driven by evolving, localized interfacial energy gradients can be a dominant mechanism, particularly on non-ideal or dynamically changing substrates.
Secondly, we provide clear evidence for such directed solid-state nanoparticle motion and assembly controlled by \textit{in situ} generated chemical heterogeneity, a phenomenon less commonly detailed than for liquid droplets or colloidal systems.
The crucial role of the Pt-to-\ce{Pt3Si} phase transformation in ``unlocking'' the necessary mobility for these solid nanoparticles to respond to interfacial gradients is a key mechanistic insight.

These results underscore the critical importance of the local nanoenvironment and interfacial energetics in predicting and controlling the stability and evolution of supported nanoparticle systems.
The ability to induce directional migration and selective coalescence through engineered substrate heterogeneity, as demonstrated here, opens potential avenues for the rational design of catalyst systems with enhanced stability or for the directed assembly of nanoscale architectures.
Future work could focus on harnessing these Si pad-induced repulsive forces and the resulting directional migration to precisely pattern or manipulate nanoparticles on surfaces, offering new strategies for the bottom-up fabrication of functional nanomaterials.

\section{Methods}\label{sec_methods}

\subsection{Synthesis of Pt Nanoparticles}

Pt NPs coated with dendrimer ligands to facilitate self-assembly at defined particle spacings were synthesized via solvothermal methods following established protocols\cite{Yang2023}. Briefly, 0.5 mmol platinum (II) acetylacetonate (Pt(acac)$_2$) was combined with 10 mL octadecene (ODE), 0.5 mL of oleylamine (OLAm) and 0.5 mL oleic acid (OAc) in a 50 mL three-neck flask. The solution was degassed on a Schlenk line under 120 °C for 15 minutes. After switching to N$_2$ atmosphere, the temperature was raised to 185 °C degrees and 10 $\mu$L of iron (0) pentacarbonyl (Fe(CO)$_5$) was injected swiftly. The reaction was maintained at 190 °C for another 1 hour until it was cooled down to room temperature. The NPs were collected by centrifugation with the addition of isopropanol (IPA) as anti-solvent and dispersed in 10 mL hexane after synthesis.

\subsection{\textit{In situ} TEM Experiments}

After synthesis, the nanoparticle solution was drop-cast onto \textit{in situ} heating chips provided by Hummingbird Scientific\cite{2024}. The chips were dried in a vacuum oven at room temperature at approximately 0.16 atm ($\approx 162.12$ hPa) overnight. Subsequently, the chips were plasma-cleaned under O$_2$ conditions (forward target power = 25 W) in two 20-second cycles to remove the ligand coating prior to heating. The cleaned chips were then loaded into a Hummingbird Scientific heating holder and inserted into a JEOL F200 S/TEM operating at 200 kV.

For real-time imaging, the samples were heated to the target experimental temperatures in 50 °C increments, and image sequences were captured every 6 seconds over a 1.5-hour period using a Gatan OneView detector with 4k pixel resolution.

During the \textit{in situ} SAED experiments, the sample was initially heated to 300 °C, followed by incremental increases of 50 °C. At each step, the temperature was held constant for 5 minutes before acquiring the SAED pattern. After reaching 800 °C, SAEDs were recorded every 5 minutes for a duration of 30 minutes. To minimize beam-induced damage, the samples were exposed to the electron beam only during the brief moments required for image acquisition.

\subsection{Image Segmentation and Analysis}

Image segmentation and particle tracking were performed using an unsupervised convolutional neural network (CNN) pipeline\cite{Horwath2020, Vyas2022}. To reduce background noise, a Gaussian filter was applied to the raw image data. A random selection of 200 experimental images from each dataset was used to construct a training set for an autoencoder model. Each image was divided into non-overlapping 512 x 512-pixel sections, and data augmentation was performed by rotating each section by 90°, 180°, and 270°. A UNet-type unsupervised CNN was retrained for each experimental dataset and applied to the filtered images using the PyTorch framework\cite{Paszke2017}. Particle sizes and positions were extracted from binary images using the connected components algorithm available in the scikit-image library\cite{VanderWalt2014}.

\subsection{Simulation Model Technique}

The choice of simulation method is dependent on the characteristic time and length scales involved in the experiment. Here nanoparticles are migrating on the order of seconds. Traditionally, molecular dynamic approaches would be utilized for nanoparticle coalescence. Such approaches have high computational cost and are limited by phononic timescales. As an alternative, we use phase field models that are instead linked to the diffusion-scales. However, due to the small length-scales involved care must be taken in how the phase field methodology is constructed.

We restrict our attention to a substrate-adhered nanoparticle coexisting with a near-vacuum environment irrespective of the material. Energy can be stored in both the short-ranged structural bonding and in long-ranged interaction potentials. The former would {\it e.g.} denote the metallic bonding that accounts for solid-state rigidity and differentiates a solid phase from a disordered counterpart (liquid/vapour). The substrate and nanoparticles both have such interactions, which we characterize through the fields $\psi(r)$ and $\phi(r)$, respectively. Each take on a value of $1$ in the solid phase and $0$ in the vacuum while the interface takes on values continuously in between. The bonding characteristics and thermodynamics are slaved through mass motion kinetics. As such, a density fields, $\rho_{\phi}(r)$ and $\rho_{\psi}(r)$, are necessitated for each dominant species. When the local density is increased, a corresponding increase in thermodynamic pressure occurs. Other non-local interactions, {\it e.g.} Van der Waals forces, can be incorporated through convolutions and gradient interactions.


The power of phase field methods lies on the description of kinetics far from equilibrium. The density and structural order evolve through gradients in the free energy, $F$, according to,
\begin{subequations}
    \begin{align}
          \pdv{\rho}{t} &= \nabla \cdot \left(M_{\rho}(\rho, \phi)\nabla\fdv{F}{\rho}\right), \label{eqn_dyn_rho}\\
          \pdv{\phi}{t} &= -M_{\phi}(\rho, \phi)\fdv{F}{\phi}.  \label{eqn_dyn_phi}
    \end{align}
\end{subequations}
The difference in the forms arises due to the requirement of total density conservation. To take into account varying diffusion modes the mobilities, $M_{i}$, are modified to account for differences in fast degrees of freedom that are coarse-grained away. Although rigorous coarse-graining results in fluctuation terms, we subsume such effects into the effective mobility coefficients. We employ an interpolated mobility,
\begin{subequations}
    \begin{align}
    M_{\rho}(\rho, \phi) & = 16\overline{M_{\rho}}\phi^2 (1 - \phi)^2 + \sum_{i} h_{i}(\phi) M_{\text{i,bulk}}(\rho),\label{eqn_Density_Mobility} \\
    M_{\phi}(\rho, \phi) & = \overline{M_\phi}.
\label{eqn_Order_Parameter_Mobility}
    \end{align}
\end{subequations}
Herein, $M_{\rho}$ can be related to surface mobility, while $M_{\text{i,bulk}}$ controls the bulk mobility of phase $i$. In general, additional mobilities are required to account for surface diffusion along the substrate. The interpolation functions have the form,
\begin{equation}\label{eqn_interpolation_function}
    h_{i}(\phi) = \frac{1}{2}\left(1 \pm \tanh\left(\frac{\phi - 0.5}{\sigma}\right)\right),
\end{equation}
with further elaboration provided in Ref.~\cite{Burns2025}. Although an additional set of evolution equations is required to account for the substrate, we make the assumption that the substrate is non-evolving. In which case the density and order do not evolve and a single representative field $\psi(r)$ denotes the substrate surface with an interface associated to interactions.

The free energy guiding the evolution is split into three main contributions:
\begin{equation}\label{eqn_free_energy}
F = F_\text{bulk} + F_{\text{gradient}} + F_{\text{barrier}}.
\end{equation}
The first accounts for the free energy in each bulk phase. This coefficient is related to the thermodynamics of each phase/species separately. Using the same interpolation function as before,
\begin{equation}\label{eqn_fe_bulk}
    F_\text{bulk} = \int_{\Omega} \sum_{i,j} h_i(\phi) h_j(\psi)f_{i,j}(\rho, \psi),
\end{equation}
where the bulk free energies are expanded quadratically with cross couplings. In particular, $f_{i,j}(\rho, \psi) = f_i(\rho) + \tilde{f}_{i,j}(\rho, \psi)$, where $f_{i} = a_{0,i} + a_{1,i}(\rho - \rho^i_{\text{ref}}) + a_{2,i}(\rho - \rho^i_{\text{ref}})^2$ is the pure material free energy. The cross-couplings are also expanded as quadratics, $\tilde{f}_{i,j} = a^{ij}_{00} + a^{ij}_{12}\rho\psi^2 + a^{ij}_{21} \rho^2 \psi + a^{ij}_{22}\rho^2 \psi^2$. 
The second contribution assigns an energy scale for interfaces and is expanded in powers of gradients:
\begin{equation}\label{eqn_fe_gradients}
    F_\text{gradient} = \int_{\Omega} \frac{|\epsilon_{\phi}\nabla \phi|^2}{2} + \frac{|\epsilon_{\rho}\nabla \rho|^2}{2} +\frac{|\epsilon_{\psi}\nabla \psi|^2}{2} + \epsilon (\nabla\phi \cdot \nabla\psi).  
\end{equation}
Here $\epsilon_{phi}$ is related to the solid-vacuum surface tension. Meanwhile, $\epsilon_{\rho}$ is associated with the long-wavelength direct correlation function. This amounts to describing pairwise interaction potentials such as van der Waals forces. The coefficient $\epsilon_{psi}$ sets the substrate-vapor interface energy. Since the substrate field is held fixed for the purposes of this article, $\epsilon_{\psi}$ does not enter the dynamics. $\epsilon$ controls the interface energy difference between substrate-vapor and substrate-solid. 

The final ingredient in the free energy is a multi-well function in terms of the order parameters to ascribe barrier heights between coexisting phases:
\begin{equation}\label{eqn_fe_barrier}
    F = \int_{\Omega} (a_0 + a_1 \psi^2) \phi^2 (1 - \phi)^2 + (b_0 + b_1 \phi^2)\psi^2 (1-\psi)^2 + c_0\psi^2 \phi^2,
\end{equation}
where the coefficients affect the order parameter interface width \cite{Burns2025}. 

The phase field model is simulated with numeric and thermodynamic parameters as listed in tables 1 and 2 in the supplementary information. An implicit time solver is utilized through the SUNDials package \cite{Hindmarsh2005,Gardner2022}. A tolerance coefficient, $r_{\text{tol}} = 10^{-5}$, was selected which ensured accuracy when perturbed. Further details of the numeric details are discussed in Ref.~\cite{Burns2025}.


\section{Acknowledgments}

This study was supported by the National Science Foundation, Division of Materials Research's Metals and Metallic Nanostructures program, DMR-2303084. This work used resources of the Singh Center for Nanotechnology, which is supported by the NSF National Nanotechnology Coordinated Infrastructure Program under grant NNCI-2025608 and through the use of facilities supported by the University of Pennsylvania Materials Research Science and Engineering Center (MRSEC) DMR-2309043. We would like to thank Lucy Decker, Shengshong Yang and Christopher Murray from the University of Pennsylvania Department of Chemistry for help with nanoparticle synthesis and self-assembly.

\newpage
\bibliography{ref_20250511}

\providecommand{\latin}[1]{#1}
\makeatletter
\providecommand{\doi}
  {\begingroup\let\do\@makeother\dospecials
  \catcode`\{=1 \catcode`\}=2 \doi@aux}
\providecommand{\doi@aux}[1]{\endgroup\texttt{#1}}
\makeatother
\providecommand*\mcitethebibliography{\thebibliography}
\csname @ifundefined\endcsname{endmcitethebibliography}  {\let\endmcitethebibliography\endthebibliography}{}
\begin{mcitethebibliography}{45}
\providecommand*\natexlab[1]{#1}
\providecommand*\mciteSetBstSublistMode[1]{}
\providecommand*\mciteSetBstMaxWidthForm[2]{}
\providecommand*\mciteBstWouldAddEndPuncttrue
  {\def\EndOfBibitem{\unskip.}}
\providecommand*\mciteBstWouldAddEndPunctfalse
  {\let\EndOfBibitem\relax}
\providecommand*\mciteSetBstMidEndSepPunct[3]{}
\providecommand*\mciteSetBstSublistLabelBeginEnd[3]{}
\providecommand*\EndOfBibitem{}
\mciteSetBstSublistMode{f}
\mciteSetBstMaxWidthForm{subitem}{(\alph{mcitesubitemcount})}
\mciteSetBstSublistLabelBeginEnd
  {\mcitemaxwidthsubitemform\space}
  {\relax}
  {\relax}

\bibitem[Cuenya(2010)]{Cuenya2010}
Cuenya,~B.~R. Synthesis and catalytic properties of metal nanoparticles: Size, shape, support, composition, and oxidation state effects. \emph{Thin Solid Films} \textbf{2010}, \emph{518}, 3127--3150\relax
\mciteBstWouldAddEndPuncttrue
\mciteSetBstMidEndSepPunct{\mcitedefaultmidpunct}
{\mcitedefaultendpunct}{\mcitedefaultseppunct}\relax
\EndOfBibitem
\bibitem[Liu and Corma(2018)Liu, and Corma]{Liu2018}
Liu,~L.; Corma,~A. Metal catalysts for heterogeneous catalysis: from single atoms to nanoclusters and nanoparticles. \emph{Chemical reviews} \textbf{2018}, \emph{118}, 4981--5079\relax
\mciteBstWouldAddEndPuncttrue
\mciteSetBstMidEndSepPunct{\mcitedefaultmidpunct}
{\mcitedefaultendpunct}{\mcitedefaultseppunct}\relax
\EndOfBibitem
\bibitem[Hansen \latin{et~al.}(2013)Hansen, DeLaRiva, Challa, and Datye]{Hansen2013}
Hansen,~T.~W.; DeLaRiva,~A.~T.; Challa,~S.~R.; Datye,~A.~K. Sintering of catalytic nanoparticles: particle migration or Ostwald ripening? \emph{Accounts of chemical research} \textbf{2013}, \emph{46}, 1720--1730\relax
\mciteBstWouldAddEndPuncttrue
\mciteSetBstMidEndSepPunct{\mcitedefaultmidpunct}
{\mcitedefaultendpunct}{\mcitedefaultseppunct}\relax
\EndOfBibitem
\bibitem[Luo \latin{et~al.}(2022)Luo, Zhao, Pan, and Sun]{Luo2022}
Luo,~Z.; Zhao,~G.; Pan,~H.; Sun,~W. Strong metal--support interaction in heterogeneous catalysts. \emph{Advanced Energy Materials} \textbf{2022}, \emph{12}, 2201395\relax
\mciteBstWouldAddEndPuncttrue
\mciteSetBstMidEndSepPunct{\mcitedefaultmidpunct}
{\mcitedefaultendpunct}{\mcitedefaultseppunct}\relax
\EndOfBibitem
\bibitem[Kim \latin{et~al.}(2010)Kim, Pint, Amama, Hauge, Maruyama, and Stach]{Kim2010}
Kim,~S.~M.; Pint,~C.~L.; Amama,~P.~B.; Hauge,~R.~H.; Maruyama,~B.; Stach,~E.~A. Catalyst and catalyst support morphology evolution in single-walled carbon nanotube supergrowth: Growth deceleration and termination. \emph{Journal of Materials Research} \textbf{2010}, \emph{25}, 1875--1885\relax
\mciteBstWouldAddEndPuncttrue
\mciteSetBstMidEndSepPunct{\mcitedefaultmidpunct}
{\mcitedefaultendpunct}{\mcitedefaultseppunct}\relax
\EndOfBibitem
\bibitem[Anderson(1989)]{Anderson1989}
Anderson,~J.~L. Colloid transport by interfacial forces. \emph{Annual review of fluid mechanics} \textbf{1989}, \emph{21}, 61--99\relax
\mciteBstWouldAddEndPuncttrue
\mciteSetBstMidEndSepPunct{\mcitedefaultmidpunct}
{\mcitedefaultendpunct}{\mcitedefaultseppunct}\relax
\EndOfBibitem
\bibitem[Brochard(1989)]{Brochard1989}
Brochard,~F. Motions of droplets on solid surfaces induced by chemical or thermal gradients. \emph{langmuir} \textbf{1989}, \emph{5}, 432--438\relax
\mciteBstWouldAddEndPuncttrue
\mciteSetBstMidEndSepPunct{\mcitedefaultmidpunct}
{\mcitedefaultendpunct}{\mcitedefaultseppunct}\relax
\EndOfBibitem
\bibitem[Chaudhury and Whitesides(1992)Chaudhury, and Whitesides]{Chaudhury1992}
Chaudhury,~M.~K.; Whitesides,~G.~M. How to make water run uphill. \emph{Science} \textbf{1992}, \emph{256}, 1539--1541\relax
\mciteBstWouldAddEndPuncttrue
\mciteSetBstMidEndSepPunct{\mcitedefaultmidpunct}
{\mcitedefaultendpunct}{\mcitedefaultseppunct}\relax
\EndOfBibitem
\bibitem[Liu \latin{et~al.}(2017)Liu, Sun, Li, Xiang, Che, Wang, and Zhou]{Liu2017}
Liu,~C.; Sun,~J.; Li,~J.; Xiang,~C.; Che,~L.; Wang,~Z.; Zhou,~X. Long-range spontaneous droplet self-propulsion on wettability gradient surfaces. \emph{Scientific reports} \textbf{2017}, \emph{7}, 7552\relax
\mciteBstWouldAddEndPuncttrue
\mciteSetBstMidEndSepPunct{\mcitedefaultmidpunct}
{\mcitedefaultendpunct}{\mcitedefaultseppunct}\relax
\EndOfBibitem
\bibitem[Sun \latin{et~al.}(2005)Sun, Feng, Gao, and Jiang]{Sun2005}
Sun,~T.; Feng,~L.; Gao,~X.; Jiang,~L. Bioinspired surfaces with special wettability. \emph{Accounts of chemical research} \textbf{2005}, \emph{38}, 644--652\relax
\mciteBstWouldAddEndPuncttrue
\mciteSetBstMidEndSepPunct{\mcitedefaultmidpunct}
{\mcitedefaultendpunct}{\mcitedefaultseppunct}\relax
\EndOfBibitem
\bibitem[Dai \latin{et~al.}(2020)Dai, Dong, and Jiang]{Dai2020}
Dai,~H.; Dong,~Z.; Jiang,~L. Directional liquid dynamics of interfaces with superwettability. \emph{Science Advances} \textbf{2020}, \emph{6}, eabb5528\relax
\mciteBstWouldAddEndPuncttrue
\mciteSetBstMidEndSepPunct{\mcitedefaultmidpunct}
{\mcitedefaultendpunct}{\mcitedefaultseppunct}\relax
\EndOfBibitem
\bibitem[Gulfam and Chen(2022)Gulfam, and Chen]{Gulfam2022}
Gulfam,~R.; Chen,~Y. Recent growth of wettability gradient surfaces: A review. \emph{Research} \textbf{2022}, \relax
\mciteBstWouldAddEndPunctfalse
\mciteSetBstMidEndSepPunct{\mcitedefaultmidpunct}
{}{\mcitedefaultseppunct}\relax
\EndOfBibitem
\bibitem[Zhang \latin{et~al.}(2022)Zhang, Barnett, and Voorhees]{zhang2022migration}
Zhang,~Q.; Barnett,~S.; Voorhees,~P. Migration of inclusions in a matrix due to a spatially varying interface energy. \emph{Scripta Materialia} \textbf{2022}, \emph{206}, 114235\relax
\mciteBstWouldAddEndPuncttrue
\mciteSetBstMidEndSepPunct{\mcitedefaultmidpunct}
{\mcitedefaultendpunct}{\mcitedefaultseppunct}\relax
\EndOfBibitem
\bibitem[Horwath \latin{et~al.}(2021)Horwath, Voorhees, and Stach]{Horwath2021}
Horwath,~J.~P.; Voorhees,~P.~W.; Stach,~E.~A. Quantifying competitive degradation processes in supported nanocatalyst systems. \emph{Nano letters} \textbf{2021}, \emph{21}, 5324--5329\relax
\mciteBstWouldAddEndPuncttrue
\mciteSetBstMidEndSepPunct{\mcitedefaultmidpunct}
{\mcitedefaultendpunct}{\mcitedefaultseppunct}\relax
\EndOfBibitem
\bibitem[Yang \latin{et~al.}(2023)Yang, LaCour, Cai, Xu, Rosen, Zhang, Kagan, Glotzer, and Murray]{Yang2023}
Yang,~S.; LaCour,~R.~A.; Cai,~Y.-Y.; Xu,~J.; Rosen,~D.~J.; Zhang,~Y.; Kagan,~C.~R.; Glotzer,~S.~C.; Murray,~C.~B. Self-assembly of atomically aligned nanoparticle superlattices from Pt–Fe3O4 heterodimer nanoparticles. \emph{Journal of the American Chemical Society} \textbf{2023}, \emph{145}, 6280--6288\relax
\mciteBstWouldAddEndPuncttrue
\mciteSetBstMidEndSepPunct{\mcitedefaultmidpunct}
{\mcitedefaultendpunct}{\mcitedefaultseppunct}\relax
\EndOfBibitem
\bibitem[202(2024)]{2024}
MEMS Heating and Biasing; Hummingbird Scientific. 2024; \url{https://hummingbirdscientific.com/products/heating-biasing/}\relax
\mciteBstWouldAddEndPuncttrue
\mciteSetBstMidEndSepPunct{\mcitedefaultmidpunct}
{\mcitedefaultendpunct}{\mcitedefaultseppunct}\relax
\EndOfBibitem
\bibitem[Ristau \latin{et~al.}(2009)Ristau, Tiruvalam, Clasen, Gorskowski, Harmer, Kiely, Hussain, and Brust]{Ristau2009}
Ristau,~R.; Tiruvalam,~R.; Clasen,~P.~L.; Gorskowski,~E.~P.; Harmer,~M.~P.; Kiely,~C.~J.; Hussain,~I.; Brust,~M. Electron microscopy studies of the thermal stability of gold nanoparticle arrays. \emph{Gold Bulletin} \textbf{2009}, \emph{42}, 133--143\relax
\mciteBstWouldAddEndPuncttrue
\mciteSetBstMidEndSepPunct{\mcitedefaultmidpunct}
{\mcitedefaultendpunct}{\mcitedefaultseppunct}\relax
\EndOfBibitem
\bibitem[Cortie(2013)]{Cortie2013}
Cortie,~D. Interface effects in magnetic metal/metal-oxide thin film systems. Ph.D.\ thesis, Ph. D. Thesis, 2013\relax
\mciteBstWouldAddEndPuncttrue
\mciteSetBstMidEndSepPunct{\mcitedefaultmidpunct}
{\mcitedefaultendpunct}{\mcitedefaultseppunct}\relax
\EndOfBibitem
\bibitem[Gohle and Schubert(1964)Gohle, and Schubert]{Gohle1964}
Gohle,~R.; Schubert,~K. On the system Platinum-Silicon. \emph{ZEITSCHRIFT FUR METALLKUNDE} \textbf{1964}, \emph{55}, 503--511\relax
\mciteBstWouldAddEndPuncttrue
\mciteSetBstMidEndSepPunct{\mcitedefaultmidpunct}
{\mcitedefaultendpunct}{\mcitedefaultseppunct}\relax
\EndOfBibitem
\bibitem[Ram and Bhan(1978)Ram, and Bhan]{Ram1978}
Ram,~R.~P.; Bhan,~S. On the constitution of platinum-silicon alloys. \emph{International Journal of Materials Research} \textbf{1978}, \emph{69}, 524--529\relax
\mciteBstWouldAddEndPuncttrue
\mciteSetBstMidEndSepPunct{\mcitedefaultmidpunct}
{\mcitedefaultendpunct}{\mcitedefaultseppunct}\relax
\EndOfBibitem
\bibitem[Lamber and Romanowski(1987)Lamber, and Romanowski]{Lamber1987}
Lamber,~R.; Romanowski,~W. Dispersion changes of platinum supported on silica glass during thermal treatment in oxygen and hydrogen atmospheres. \emph{Journal of Catalysis} \textbf{1987}, \emph{105}, 213--226\relax
\mciteBstWouldAddEndPuncttrue
\mciteSetBstMidEndSepPunct{\mcitedefaultmidpunct}
{\mcitedefaultendpunct}{\mcitedefaultseppunct}\relax
\EndOfBibitem
\bibitem[Wang \latin{et~al.}(2003)Wang, Penner, Su, Rupprechter, Hayek, and Schlögl]{Wang2003}
Wang,~D.; Penner,~S.; Su,~D.~S.; Rupprechter,~G.; Hayek,~K.; Schlögl,~R. Silicide formation on a Pt/SiO2 model catalyst studied by TEM, EELS, and EDXS. \emph{Journal of Catalysis} \textbf{2003}, \emph{219}, 434--441\relax
\mciteBstWouldAddEndPuncttrue
\mciteSetBstMidEndSepPunct{\mcitedefaultmidpunct}
{\mcitedefaultendpunct}{\mcitedefaultseppunct}\relax
\EndOfBibitem
\bibitem[Behafarid \latin{et~al.}(2014)Behafarid, Pandey, Diaz, Stach, and Cuenya]{Behafarid2014}
Behafarid,~F.; Pandey,~S.; Diaz,~R.~E.; Stach,~E.~A.; Cuenya,~B.~R. An in situ transmission electron microscopy study of sintering and redispersion phenomena over size-selected metal nanoparticles: environmental effects. \emph{Physical Chemistry Chemical Physics} \textbf{2014}, \emph{16}, 18176--18184\relax
\mciteBstWouldAddEndPuncttrue
\mciteSetBstMidEndSepPunct{\mcitedefaultmidpunct}
{\mcitedefaultendpunct}{\mcitedefaultseppunct}\relax
\EndOfBibitem
\bibitem[Lamber and Jaeger(1991)Lamber, and Jaeger]{Lamber1991}
Lamber,~R.; Jaeger,~N.~I. On the reaction of Pt with SiO2 substrates: Observation of the Pt3Si phase with the Cu3Au superstructure. \emph{Journal of applied physics} \textbf{1991}, \emph{70}, 457--461\relax
\mciteBstWouldAddEndPuncttrue
\mciteSetBstMidEndSepPunct{\mcitedefaultmidpunct}
{\mcitedefaultendpunct}{\mcitedefaultseppunct}\relax
\EndOfBibitem
\bibitem[Canali \latin{et~al.}(1977)Canali, Catellani, Prudenziati, Wadlin, and Evans~Jr]{Canali1977}
Canali,~C.; Catellani,~C.; Prudenziati,~M.; Wadlin,~W.; Evans~Jr,~C. Pt2Si and PtSi formation with high‐purity Pt thin films. \emph{Applied Physics Letters} \textbf{1977}, \emph{31}, 43--45\relax
\mciteBstWouldAddEndPuncttrue
\mciteSetBstMidEndSepPunct{\mcitedefaultmidpunct}
{\mcitedefaultendpunct}{\mcitedefaultseppunct}\relax
\EndOfBibitem
\bibitem[Canali \latin{et~al.}(1979)Canali, Majni, Ottaviani, and Celotti]{Canali1979}
Canali,~C.; Majni,~G.; Ottaviani,~G.; Celotti,~G. Phase diagrams and metal‐rich silicide formation. \emph{Journal of Applied Physics} \textbf{1979}, \emph{50}, 255--258\relax
\mciteBstWouldAddEndPuncttrue
\mciteSetBstMidEndSepPunct{\mcitedefaultmidpunct}
{\mcitedefaultendpunct}{\mcitedefaultseppunct}\relax
\EndOfBibitem
\bibitem[Ottaviani and Costato(1978)Ottaviani, and Costato]{Ottaviani1978}
Ottaviani,~G.; Costato,~M. Compound formation in metal—semiconductor interactions. \emph{Journal of Crystal Growth} \textbf{1978}, \emph{45}, 365--375\relax
\mciteBstWouldAddEndPuncttrue
\mciteSetBstMidEndSepPunct{\mcitedefaultmidpunct}
{\mcitedefaultendpunct}{\mcitedefaultseppunct}\relax
\EndOfBibitem
\bibitem[Horwath \latin{et~al.}(2020)Horwath, Zakharov, Mégret, and Stach]{Horwath2020}
Horwath,~J.~P.; Zakharov,~D.~N.; Mégret,~R.; Stach,~E.~A. Understanding important features of deep learning models for segmentation of high-resolution transmission electron microscopy images. \emph{npj Computational Materials} \textbf{2020}, \emph{6}, 108\relax
\mciteBstWouldAddEndPuncttrue
\mciteSetBstMidEndSepPunct{\mcitedefaultmidpunct}
{\mcitedefaultendpunct}{\mcitedefaultseppunct}\relax
\EndOfBibitem
\bibitem[Leena~Vyas(2022)]{Vyas2022}
Leena~Vyas,~E. A.~S.,~James P.~Horwath Tutorial on Unsupervised Image Segmentation for Electron Microscopy. 2022; \url{https://mlforem.github.io}\relax
\mciteBstWouldAddEndPuncttrue
\mciteSetBstMidEndSepPunct{\mcitedefaultmidpunct}
{\mcitedefaultendpunct}{\mcitedefaultseppunct}\relax
\EndOfBibitem
\bibitem[Jeon \latin{et~al.}(2021)Jeon, Heo, Hwang, Ciston, Bustillo, Reed, Ham, Kang, Kim, and Lim]{Jeon2021}
Jeon,~S.; Heo,~T.; Hwang,~S.-Y.; Ciston,~J.; Bustillo,~K.~C.; Reed,~B.~W.; Ham,~J.; Kang,~S.; Kim,~S.; Lim,~J. Reversible disorder-order transitions in atomic crystal nucleation. \emph{Science} \textbf{2021}, \emph{371}, 498--503\relax
\mciteBstWouldAddEndPuncttrue
\mciteSetBstMidEndSepPunct{\mcitedefaultmidpunct}
{\mcitedefaultendpunct}{\mcitedefaultseppunct}\relax
\EndOfBibitem
\bibitem[Meng \latin{et~al.}(2021)Meng, Low, Foucher, Li, Petrovic, and Stach]{meng2021anomalous}
Meng,~A.~C.; Low,~K.-B.; Foucher,~A.~C.; Li,~Y.; Petrovic,~I.; Stach,~E.~A. Anomalous metal vaporization from Pt/Pd/Al 2 O 3 under redox conditions. \emph{Nanoscale} \textbf{2021}, \emph{13}, 11427--11438\relax
\mciteBstWouldAddEndPuncttrue
\mciteSetBstMidEndSepPunct{\mcitedefaultmidpunct}
{\mcitedefaultendpunct}{\mcitedefaultseppunct}\relax
\EndOfBibitem
\bibitem[Roth(1994)]{Roth1994}
Roth,~A. \emph{Vacuum sealing techniques}; Springer Science \& Business Media, 1994\relax
\mciteBstWouldAddEndPuncttrue
\mciteSetBstMidEndSepPunct{\mcitedefaultmidpunct}
{\mcitedefaultendpunct}{\mcitedefaultseppunct}\relax
\EndOfBibitem
\bibitem[Lin \latin{et~al.}(2015)Lin, Chen, Su, Wu, Hsiao, Shiao, and Chang]{Lin2015}
Lin,~C.-T.; Chen,~Y.-W.; Su,~J.; Wu,~C.-T.; Hsiao,~C.-N.; Shiao,~M.-H.; Chang,~M.-N. Facile Preparation of a Platinum Silicide Nanoparticle-Modified Tip Apex for Scanning Kelvin Probe Microscopy. \emph{Nanoscale Research Letters} \textbf{2015}, \emph{10}, 1--6\relax
\mciteBstWouldAddEndPuncttrue
\mciteSetBstMidEndSepPunct{\mcitedefaultmidpunct}
{\mcitedefaultendpunct}{\mcitedefaultseppunct}\relax
\EndOfBibitem
\bibitem[Wu \latin{et~al.}(2015)Wu, Wang, Wu, Yang, Liu, Han, Xu, and Wang]{Wu2015}
Wu,~D.; Wang,~P.; Wu,~P.; Yang,~Q.; Liu,~F.; Han,~Y.; Xu,~F.; Wang,~L. Determination of contact angle of droplet on convex and concave spherical surfaces. \emph{Chemical Physics} \textbf{2015}, \emph{457}, 63--69\relax
\mciteBstWouldAddEndPuncttrue
\mciteSetBstMidEndSepPunct{\mcitedefaultmidpunct}
{\mcitedefaultendpunct}{\mcitedefaultseppunct}\relax
\EndOfBibitem
\bibitem[Hara \latin{et~al.}(2005)Hara, Izumi, Kumagai, and Sakai]{Hara2005}
Hara,~S.; Izumi,~S.; Kumagai,~T.; Sakai,~S. Surface energy, stress and structure of well-relaxed amorphous silicon: A combination approach of ab initio and classical molecular dynamics. \emph{Surface Science} \textbf{2005}, \emph{585}, 17--24\relax
\mciteBstWouldAddEndPuncttrue
\mciteSetBstMidEndSepPunct{\mcitedefaultmidpunct}
{\mcitedefaultendpunct}{\mcitedefaultseppunct}\relax
\EndOfBibitem
\bibitem[Lu \latin{et~al.}(2005)Lu, Huang, Cuma, and Liu]{Lu2005}
Lu,~G.-H.; Huang,~M.; Cuma,~M.; Liu,~F. Relative stability of Si surfaces: A first-principles study. \emph{Surface Science} \textbf{2005}, \emph{588}, 61--70\relax
\mciteBstWouldAddEndPuncttrue
\mciteSetBstMidEndSepPunct{\mcitedefaultmidpunct}
{\mcitedefaultendpunct}{\mcitedefaultseppunct}\relax
\EndOfBibitem
\bibitem[Kim \latin{et~al.}(2018)Kim, Seol, and Lee]{Kim2018}
Kim,~J.-S.; Seol,~D.; Lee,~B.-J. Critical assessment of Pt surface energy -- An atomistic study. \emph{Surface Science} \textbf{2018}, \emph{670}, 8--12\relax
\mciteBstWouldAddEndPuncttrue
\mciteSetBstMidEndSepPunct{\mcitedefaultmidpunct}
{\mcitedefaultendpunct}{\mcitedefaultseppunct}\relax
\EndOfBibitem
\bibitem[Hamaker(1937)]{Hamaker1937}
Hamaker,~H. The London---van der Waals attraction between spherical particles. \emph{Physica} \textbf{1937}, \emph{4}, 1058--1072\relax
\mciteBstWouldAddEndPuncttrue
\mciteSetBstMidEndSepPunct{\mcitedefaultmidpunct}
{\mcitedefaultendpunct}{\mcitedefaultseppunct}\relax
\EndOfBibitem
\bibitem[Tolias(2020)]{Tolias2020}
Tolias,~P. Non-retarded room temperature Hamaker constants between elemental metals. \emph{Surface Science} \textbf{2020}, \emph{700}, 121652\relax
\mciteBstWouldAddEndPuncttrue
\mciteSetBstMidEndSepPunct{\mcitedefaultmidpunct}
{\mcitedefaultendpunct}{\mcitedefaultseppunct}\relax
\EndOfBibitem
\bibitem[Paszke \latin{et~al.}(2017)Paszke, Gross, Chintala, Chanan, Yang, DeVito, Lin, Desmaison, Antiga, and Lerer]{Paszke2017}
Paszke,~A.; Gross,~S.; Chintala,~S.; Chanan,~G.; Yang,~E.; DeVito,~Z.; Lin,~Z.; Desmaison,~A.; Antiga,~L.; Lerer,~A. Automatic differentiation in pytorch. \textbf{2017}, \relax
\mciteBstWouldAddEndPunctfalse
\mciteSetBstMidEndSepPunct{\mcitedefaultmidpunct}
{}{\mcitedefaultseppunct}\relax
\EndOfBibitem
\bibitem[Van~der Walt \latin{et~al.}(2014)Van~der Walt, Schönberger, Nunez-Iglesias, Boulogne, Warner, Yager, Gouillart, and Yu]{VanderWalt2014}
Van~der Walt,~S.; Schönberger,~J.~L.; Nunez-Iglesias,~J.; Boulogne,~F.; Warner,~J.~D.; Yager,~N.; Gouillart,~E.; Yu,~T. scikit-image: image processing in Python. \emph{PeerJ} \textbf{2014}, \emph{2}, e453\relax
\mciteBstWouldAddEndPuncttrue
\mciteSetBstMidEndSepPunct{\mcitedefaultmidpunct}
{\mcitedefaultendpunct}{\mcitedefaultseppunct}\relax
\EndOfBibitem
\bibitem[Burns and Voorhees(2025)Burns, and Voorhees]{Burns2025}
Burns,~D.; Voorhees,~P. (unpublished)\relax
\mciteBstWouldAddEndPuncttrue
\mciteSetBstMidEndSepPunct{\mcitedefaultmidpunct}
{\mcitedefaultendpunct}{\mcitedefaultseppunct}\relax
\EndOfBibitem
\bibitem[Hindmarsh \latin{et~al.}(2005)Hindmarsh, Brown, Grant, Lee, Serban, Shumaker, and Woodward]{Hindmarsh2005}
Hindmarsh,~A.~C.; Brown,~P.~N.; Grant,~K.~E.; Lee,~S.~L.; Serban,~R.; Shumaker,~D.~E.; Woodward,~C.~S. {SUNDIALS}: Suite of nonlinear and differential/algebraic equation solvers. \emph{ACM Transactions on Mathematical Software (TOMS)} \textbf{2005}, \emph{31}, 363--396\relax
\mciteBstWouldAddEndPuncttrue
\mciteSetBstMidEndSepPunct{\mcitedefaultmidpunct}
{\mcitedefaultendpunct}{\mcitedefaultseppunct}\relax
\EndOfBibitem
\bibitem[Gardner \latin{et~al.}(2022)Gardner, Reynolds, Woodward, and Balos]{Gardner2022}
Gardner,~D.~J.; Reynolds,~D.~R.; Woodward,~C.~S.; Balos,~C.~J. Enabling new flexibility in the {SUNDIALS} suite of nonlinear and differential/algebraic equation solvers. \emph{ACM Transactions on Mathematical Software (TOMS)} \textbf{2022}, \relax
\mciteBstWouldAddEndPunctfalse
\mciteSetBstMidEndSepPunct{\mcitedefaultmidpunct}
{}{\mcitedefaultseppunct}\relax
\EndOfBibitem
\end{mcitethebibliography}


\end{document}